\documentclass[lettersize,journal]{IEEEtran}
\usepackage{amsmath,amsfonts}
\usepackage{algorithmic}
\usepackage{algorithm}
\usepackage{array}
\usepackage[caption=false,font=normalsize,labelfont=sf,textfont=sf]{subfig}
\usepackage{textcomp}
\usepackage{stfloats}
\usepackage{url}
\usepackage{verbatim}
\usepackage{graphicx}
\usepackage{cite}
\usepackage{multirow}
\usepackage[table,xcdraw]{xcolor}
\begin{document}

\title{Machine Anomalous Sound Detection Using Spectral-temporal Modulation Representations Derived from Machine-specific Filterbanks}

\author{Kai Li, Khalid Zaman, Xingfeng Li, Masato Akagi, and Masashi Unoki
\thanks{Kai Li, Khalid Zaman, Masato Akagi, and Masashi Unoki are with the Graduate School of Advanced Science and Technology, Japan Advanced Institute of Science and Technology, Nomi, Ishikawa
923-1292, Japan (e-mail: kai\_li@jaist.ac.jp; zaman.khalid@jaist.ac.jp;
unoki@jaist.ac.jp).}
\thanks{Xingfeng Li is with the Faculty of Data Science, City University of Macau, Macau, China (e-mail: xfli@cityu.edu.mo).}
}




\maketitle

\begin{abstract}
Early detection of factory machinery malfunctions is crucial in industrial applications. In machine anomalous sound detection (ASD), different machines exhibit unique vibration-frequency ranges based on their physical properties. Meanwhile, the human auditory system is adept at tracking both temporal and spectral dynamics of machine sounds. Consequently, integrating the computational auditory models of the human auditory system with machine-specific properties can be an effective approach to machine ASD. We first quantified the frequency importances of four types of machines using the Fisher ratio (F-ratio). The quantified frequency importances were then used to design machine-specific non-uniform filterbanks (NUFBs), which extract the log non-uniform spectrum (LNS) feature. The designed NUFBs have a narrower bandwidth and higher filter distribution density in frequency regions with relatively high F-ratios. Finally, spectral and temporal modulation representations derived from the LNS feature were proposed. These proposed LNS feature and modulation representations are input into an autoencoder neural-network-based detector for ASD. The quantification results from the training set of the Malfunctioning Industrial Machine Investigation and Inspection dataset with a signal-to-noise (SNR) of 6 dB reveal that the distinguishing information between normal and anomalous sounds of different machines is encoded non-uniformly in the frequency domain. By highlighting these important frequency regions using NUFBs, the LNS feature can significantly enhance performance using the metric of AUC (area under the receiver operating characteristic curve) under various SNR conditions. Furthermore, modulation representations can further improve performance. Specifically, temporal modulation is effective for fans, pumps, and sliders, while spectral modulation is particularly effective for valves.
\end{abstract}


\section{Introduction}
\IEEEPARstart{A}nomalous sound detection (ASD) for machine condition monitoring enables workers to arrange maintenance work to fix machine problems in the earliest stages of the anomaly, thus preventing sustained damage, reducing maintenance costs and optimizing production efficiency \cite{koizumi2019toyadmos}. Developing advanced ASD systems is an important component of the fourth industrial revolution and has received increasing attention \cite{kawaguchi2017can,nunes2021anomalous}. 

ASD can be classified into two types of problems \cite{perez2020anomalous}, that is, supervised ASD in which recordings of anomalous events to be detected are available in training, and unsupervised ASD in which recordings of the anomalous events are not available in training. Most methods for ASD are based on an unsupervised autoencoder (AE) model \cite{dohi2022description,suefusa2020anomalous,kapka2020id} because of difficulties in collecting anomalous sounds that can cover all possible types of anomalies \cite{koizumi2018unsupervised}. These methods are used to detect “\textit{unknown}” anomalous sounds that have not been observed using reconstruction errors. However, because the training procedure does not incorporate anomalous sounds, the effectiveness of such models may be limited if the learned features also fit with the anomalous sounds.

Many sophisticated models have been adapted and applied to further improve the effectiveness of back-end detectors \cite{hayashi2020conformer,li2018anomalous, zaman2023survey}. For example, the WaveNet architecture was used by Hayashi, et al. \cite{hayashi2018anomalous}. Marchi et al. \cite{marchi2017deep} proposed an ASD approach that uses a denoising AE architecture with feedforward and long-short-term-memory units. The self-supervised approaches were used \cite{morita2021anomalous,giri2020self} to provide some additional information, i.e., machine type and machine identity. Hoang et al. \cite{van2021unsupervised} proposed an AE architecture, called Fully-Connected U-Net, to replace the conventional AE model. However, the performance of deep neural network (DNN)-based ASD methods depends significantly on the discrimination of acoustic front-ends. The acoustic front-end refers to the pre-processing stage that extracts features from raw audio signals before feeding them into the DNN. If the front-end can effectively capture and represent the relevant acoustic features, the DNN will have better discrimination power. 

\begin{figure}[t]
\centering
\includegraphics[width=3.5in]{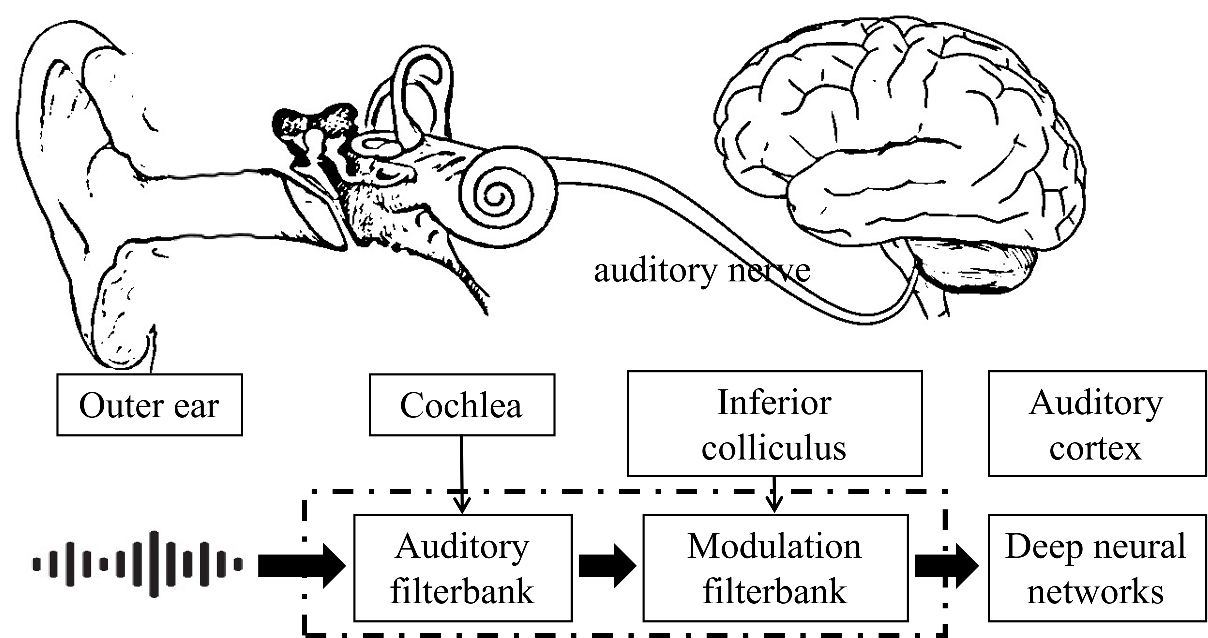}
\caption{Human auditory mechanism and its modeling.}
\label{fig:auditory}
\end{figure}

The human auditory system has been shown to be effective and robust against noise in many types of recognition tests \cite{kim1999auditory, luo2008auditory, yuncu2014automatic}. Therefore, it is intuitive to include the properties of human auditory in the extraction of acoustic front-ends for potential performance enhancement. Figure \ref{fig:auditory} illustrates the auditory processing pathway. This pathway begins in the outer ear, where sound waves are captured and directed to the cochlea. In the cochlea, these sound waves are transformed into electrical signals by hair cells. These signals are then transmitted through the auditory nerve to the inferior colliculus and further processed in the auditory cortex. The novel aspect of this paper is investigating acoustic features and representations inspired by various stages of the auditory system for machine ASD.



Various computational auditory models (CAMs) have been proposed to simulate these auditory processes. They are based on neurophysiological, biophysical, and psychoacoustical investigations at various stages of the auditory system \cite{yang1992auditory, wang1995spectral, lyon1996auditory}. CAMs generally include two stages. The first stage models the transformation of the acoustic signal into an internal neural representation in the cochlea, referred to as an auditory spectrogram. The second stage analyzes the spectrogram to estimate the content of its spectral and temporal modulations using a bank of modulation-selective filters, mimicking those described in a model of the mammalian primary auditory cortex \cite{wang1995spectral}. The outputs of the second stage are spectral-temporal multi-resolution representations, which are responsible for extracting the key features upon which the detection is based. These computational models improve our understanding of auditory perception and have practical applications in fields such as speech recognition \cite{schadler2012spectro, zahorian2009spectral}, speech intelligibility prediction \cite{elhilali2003spectro, edraki2020speech}, speaker identification \cite{chi2012spectro}, and audio quality assessment \cite{nguyen2022automatic}. However, the complexity of the human auditory mechanism makes it challenging to fully understand the audio signal processing process and determine which model best mimics this process.

In machine ASD, different types of machines have different vibration frequency regions depending on their physical property. Consequently, the discriminative information of sounds emitted from different types of machine may thus be encoded non-uniformly in the frequency domain. Auditory filterbanks, such as the Gammatone filterbank (GFB), which are often designed with a higher resolution (narrower bandwidth and higher filter distribution density) in the low-frequency regions and lower resolution (wider bandwidth and lower filter distribution density) in the high-frequency regions, may filter out important information in the high-frequency regions, decreasing the performance of an ASD system. Therefore, it is necessary to quantify the importance of frequencies and design machine-specific filterbanks, which encompass a more reasonable design in bandwidth and distribution density of filters for machine ASD. 

How can the importance of different frequencies be quantified? The Fisher ratio (F-ratio) is a statistical-based method and widely used to measure the discriminative ability of a feature for pattern recognition \cite{webb2003statistical}. It has been used to evaluate the importance of different frequencies in speaker recognition \cite{lu2008investigation,li2022relationship}, emotion recognition \cite{zhou2009physiologically, peng2020speech}, and replay attack detection \cite{hyon2014detection}. The calculation of the F-ratio does not require training data and is comparatively straightforward and efficient.

We previously proposed to quantify the importance of frequency in the ASD of sounds produced by different machines using the F-ratio, called the machine-wise F-ratio \cite{li2023data}. We were the first to quantify the frequency importance of machines using the statistically based method to detect anomalous sounds. With the quantification results, we visualized where the discriminative features of each machine are encoded in the frequency domain. To highlight such important frequency bands, we designed machine-specific non-uniform filterbanks (NUFBs) that have high resolutions in the frequency regions with high F-ratios and extract the log non-uniform spectrum (LNS) features. However, the effectiveness of modulation representations derived from such well-designed NUFBs have not been investigated in machine ASD.

Theoretically, spectral-temporal modulation (STM) representations, derived from well-designed machine-specific NUFBs, can significantly enhance the effectiveness of machine ASD. These NUFBs are tailored to capture the distinct physical characteristics of different machines, ensuring that the proposed LNS feature are highly relevant for distinguishing between normal and anomalous conditions. Spectral modulation (SM) encompasses variations in pitch and harmonic structure, which are crucial for identifying deviations in the sound profiles of machinery \cite{edraki2022spectro}. Temporal modulation (TM), on the other hand, includes features such as rhythm and timbre, which involve attributes such as sharpness, roughness, brightness, boominess, and depth \cite{ding2017temporal,ota2023anomalous}. By integrating SM and TM information, these representations provide a broad understanding of the frequency and time-varying characteristics of machine sounds. 

This study builds on our previous work \cite{li2023data} by integrating machine-specific NUFBs into the calculation of STM representations. The primary contributions of this study are as follows. First, we quantified and visualized the importance of frequency for various machines, including fans, pumps, sliders, and valves. We then designed machine-specific NUFBs and proposed LNS feature to obtain more discriminative information for machine ASD. We next proposed modulation representations, including SM, TM, and STM, derived from the LNS feature. Subsequently, we applied LNS and modulation representations to machine ASD using data with varying signal-to-noise ratios (SNRs= -6, 0, and 6 dB). Experimental results on the Malfunctioning Industrial Machine Investigation and Inspection (MIMII) database confirm the accuracy of our frequency quantification and the effectiveness of the proposed LNS feature and their modulation representations in enhancing machine ASD performance.

In Section \ref{sec:auditorymodel}, we introduce the basis of CAMs. In Section \ref{sec:proposedmethod}, we introduce the calculation of NUFBs, LNS, and their modulation representations for machine ASD. In Section \ref{sec:AEmodel}, we describe the AE model used in this study. In Section \ref{sec:data}, we provide database statistics. We give details of the experimental setup and evaluation metric in Section \ref{sec:ExpermentalSetup} and present the results and discussion in Sections \ref{sec:results} and \ref{sec:Discussion}, respectively. We conclude the paper in Section \ref{sec:conclusion}.


\begin{figure}[t]
\centering
\includegraphics[width=3.5in]{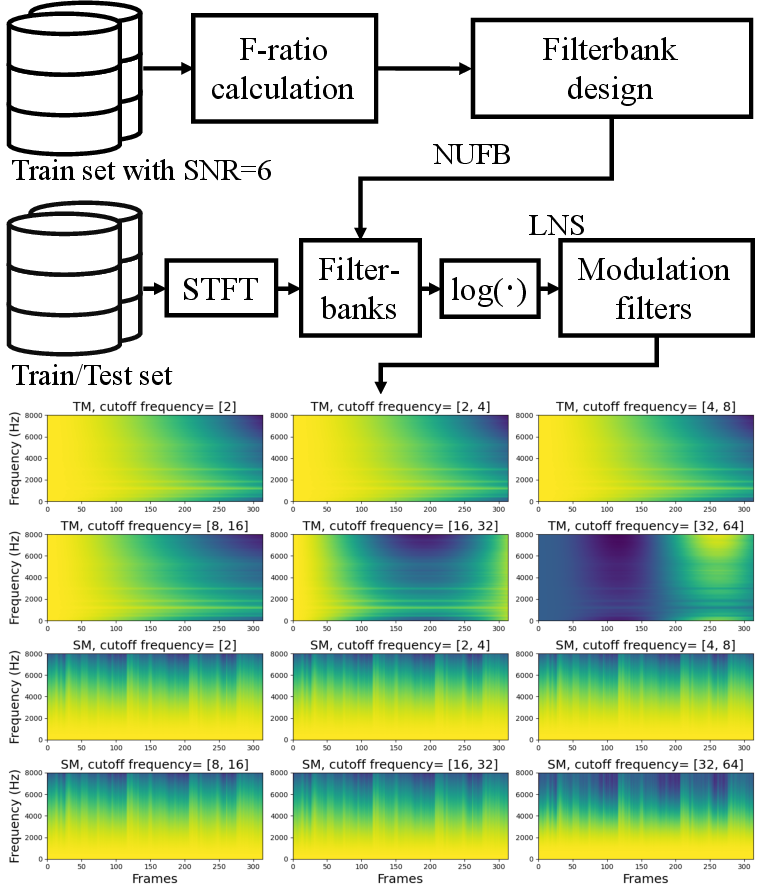}
\caption{Block diagram of spectral-temporal modulation analysis scheme.}
\label{fig:stmdiagram}
\end{figure}

\section{Computational Auditory Models}
\label{sec:auditorymodel}
To establish a foundation for the proposed features and representations, we briefly discuss the CAMs presented in previous studies \cite{mesgarani2006discrimination} and \cite{chi2012robust}. As illustrated in Fig. \ref{fig:auditory}, there are basically two modules in CAMs. The first one is the cochlea module for the auditory-spectral analysis, and the second one is the central cortical module for modulation analysis.

\subsection{Cochlea Module for Auditory-spectral Analysis}
The cochlea module models the peripheral functions of the auditory system, specifically the role of the cochlea as a frequency analyzer. Chi et al. \cite{chi2012robust} used a bank of 128 overlapping asymmetric constant-Q bandpass filters to mimic the frequency selectivity of the cochlea. These filters decompose the acoustic signal into distinct frequency components, simulating how the basilar membrane in the cochlea processes sound.

The output from each filter undergoes non-linear compression, mimicking the saturation behavior of inner hair cells, followed by a lateral inhibitory network that enhances spectral resolution by sharpening the filter responses. The envelope of the signal is then extracted, producing an auditory spectrogram that represents the time-frequency distribution of the sound. 

The auditory spectrogram produced with the linear cochlear module is similar to the magnitude response of a Mel-scaled Fourier-transform-based spectrogram. This similarity is due to the constant-Q criterion of the filterbank in the cochlear module, which sets the bandwidth proportional to the center frequency. This criterion mirrors the effects of logarithmic filter spacing in the Mel scale. Additionally, the local envelope extraction process in the cochlear module also approximates the magnitude of a Fourier transform-based spectrogram, capturing the energy distribution across frequencies over time in a manner consistent with how a Mel-scaled spectrogram operates. Therefore, the extracted auditory spectrogram can also be replaced with a Mel spectrogram through a simplified method, as in a previous study \cite{edraki2020speech}.

The Mel filterbank (MFB) and GFB are widely used and share similar characteristics, as they both have the property that the bandwidth is proportional to the center frequency. The GFB is considered the best representation of the auditory characteristics of the cochlea because it closely mimics the frequency selectivity of the human auditory system. The MFB is the most commonly used in engineering applications due to its simplicity and effectiveness. However, in machine ASD, the importance of different frequency regions may not decrease with the increase in frequency for different types of machines. These two filterbanks, particularly the MFB, may not be optimal for machine ASD because they are designed to mimic human hearing rather than capturing the specific spectral-temporal characteristics that may be more relevant for identifying anomalies in machine sounds.

\subsection{Central Cortical Module for Modulation Analysis}
The central cortical module models the spectro-temporal selectivity of neurons within the auditory cortex, focusing on how these neurons process and analyze complex auditory signals. This module further analyzes the auditory spectrogram, which is initially generated by the cochlear module, by applying two-dimensional filters tuned to various STM parameters. These parameters include $rate$, which reflects the velocity of temporal variations in the spectro-temporal envelope (measured in Hz), and $scale$, which characterizes the density of these variations along the log-frequency axis (measured in cycles per octave).

The output of this cortical processing is referred to as the STM representation, which captures the spectral-temporal structures of an input sound at each time instant. These structures encompass various auditory features, such as pitch, harmonicity, formant spacing, amplitude modulation (AM), and frequency modulation (FM). Pitch, AM, and FM, are closely related to the prosody and timbre of the sound, while the others are related to its spectral characteristics. The STM representation, therefore, encodes both prosodic and spectral features, making it a valuable tool for tasks such as speech emotion recognition \cite{peng2020speech, chi2012robust}.


\section{Proposed Methods for Machine ASD}
\label{sec:proposedmethod}
Figure \ref{fig:stmdiagram} illustrates the block diagram of the proposed STM analysis method. The process begins with the training dataset, which has an SNR of 6 dB. The first step involves calculating the F-ratio, which is then used to design machine-specific NUFBs. Subsequently, the short-time Fourier transform (STFT) is applied to the training set to obtain the time-frequency representation. The output from the STFT is processed through the designed NUFBs, and the resulting power spectrum is converted to a logarithmic scale before being passed through a modulation filterbank. The output of the logarithmic operation is the LNS feature. This sequence produces two sets of modulation representations: TM and SM, as illustrated in the lower part of the figure. This section focuses on the three main components of the STM analysis process: the calculation of the machine-wise F-ratio, design of machine-specific NUFBs, and STM analysis.

\subsection{Calculation of Machine-wise F-ratio}
The frequency bands with more discriminative features should have high inter-class variances and low intra-class variances between normal and anomalous sound classes~\cite{lu2008investigation}. Therefore, we define the F-ratio for machine $m$ as
\begin{equation}
{\rm{\mbox{$F_m$}}}=\frac{\frac{1}{2}\sum_c(u_{m,c}-u_{m})^{2}}{\frac{1}{2N}\sum_c\sum_{i=1}^N(x_{m,c}^i-u_{m,c})^{2}},
\label{fratio}
\end{equation}
where \(x_{m,c}^i\) is the sub-band energy of the $i$-th audio of class $c$ with $i=1,2,...,N$, $m\in\{\text{fan}, \text{pump}, \text{slider}, \text{valve}\}$, and $c\in\{\text{normal}, \text{anomaly}\}$. $u_{m,c}$ and $u_m$ are defined as:
\begin{equation}
    u_{m,c}=\frac{1}{N}\sum_{i=1}^Nx_{m,c}^i \qquad
\end{equation}
\begin{equation}
     \qquad u_m=\frac{1}{2N}\sum_c\sum_{i=1}^Nx_{m,c}^i
\end{equation}
They are used to calculate the variables that represent the sub-band energy averages for class $c$ and for all classes, respectively.

Equation (\ref{fratio}) is the ratio between the inter-class and the intra-class variances of the speech power in a given frequency band. A larger value obtained in a frequency band means that more discriminative information is encoded in that band.

\subsection{Design of Machine-specific Non-uniform Filterbanks}
To show the correctness of the quantification results, we designed NUFBs and used them to extract the LNS feature for each machine. The NUFBs were designed by highlighting (narrower bandwidth and higher filter distribution density) the frequency bands with relatively high F-ratios. 

The distribution density of the triangular band-pass filters is assigned to be directly proportional to the F-ratios. On the contrary, the bandwidth of the triangular band-pass filters is assigned to be inversely proportional to the F-ratios. The steps for designing an NUFB are as follows:  
\begin{enumerate}{}{}
\item{calculate the weight $k$ on the basis of the $F_m$, \(k=f_s/{(2\times\sum{F_m})}\), where $f_s$ is the sampling frequency,} 
\item {calculate the cumulative sum of the weighted $F_m$, $\text{CS}=\text{Cumsum}(k\times F_m)$,} 
\item {fit the curve of the mapping frequency from the linear scale to the adaptive scale by using the cubic spline interpolation,} 
\item {calculate the center frequencies and bandwidth of the triangular band-pass filters $C(j)$ on the basis of the fitting curve, and} 
\item {design an NUFB with the non-uniform resolutions.}
\end{enumerate}

Finally, the LNS feature are extracted by replacing the filterbank used in the extraction processes of the log Mel spectrum (LMS) and log Gammatone spectrum (LGS) features. 


\subsection{Spectral-temporal Modulation Analysis}
This section details the extraction of SM, TM, and STM representations using modulation filters with specific cut-off frequencies and designed machine-specific NUFBs. Let us define the output of the filterbank design as $s(n,t)$, where $n$ refers to the number of channels in the filterbanks, and t is the number of frames. The $k$th modulated representation in the $n$th channel can then be calculated as

\begin{equation}
    s(n, k, t) = r(k, t) * s(n, t), 1\leq k \leq K,
\end{equation}
where $r(k, t)$ is the impulse response of the $k$th modulation filter, and $K$ is the number of modulation channels. Different modulation channels have different cut-off frequency. A high modulation frequency corresponds to fast modulations and vice versa.

SM focuses on capturing variations in the spectral content of the sound signal. It provides insights into how the energy distribution across different frequency bands changes over time. TM captures the temporal dynamics of the sound signal, emphasizing how the amplitude varies over time. This is particularly useful for identifying temporal patterns such as periodicity or irregularities in machine sounds. STM combines both spectral and temporal information, providing a comprehensive representation of the modulation characteristics of audio. These multi-resolution feature representations capture the interaction between spectral and temporal variations, providing a robust tool for distinguishing between normal and anomalous sounds.

\section{Anomaly Sound Detection Using Autoencoder Model}
\label{sec:AEmodel}
The AE model consists of three primary modules: the encoder, bottleneck layer, and decoder. The encoder compresses the input data into a lower-dimensional representation, capturing the essential features of the normal sound. The bottleneck layer, which has the smallest number of neurons, acts as a constraint that forces the model to learn the most important features. The decoder then reconstructs the input data from this compressed representation.

Each module of the AE model is constructed using fully connected layers. The batch normalization and activation functions of the rectified linear unit (ReLU) are followed after each fully connected layer. The mean squared error (MSE) is utilized as the cost function to optimize the AE model. The MSE measures the average squared difference between the input and the reconstructed output, providing a clear indication of how well the model has learned to replicate normal sounds. 

During training, the AE model learns to minimize the MSE for normal sound examples, effectively capturing their inherent patterns and characteristics. In the testing phase, sound samples are passed through the trained AE model, and the reconstruction error is calculated. Sound samples that result in high reconstruction errors are flagged as anomalous. 


\begin{figure*}[t]
\centering
\includegraphics[width=\textwidth,trim=100 0 100 0]{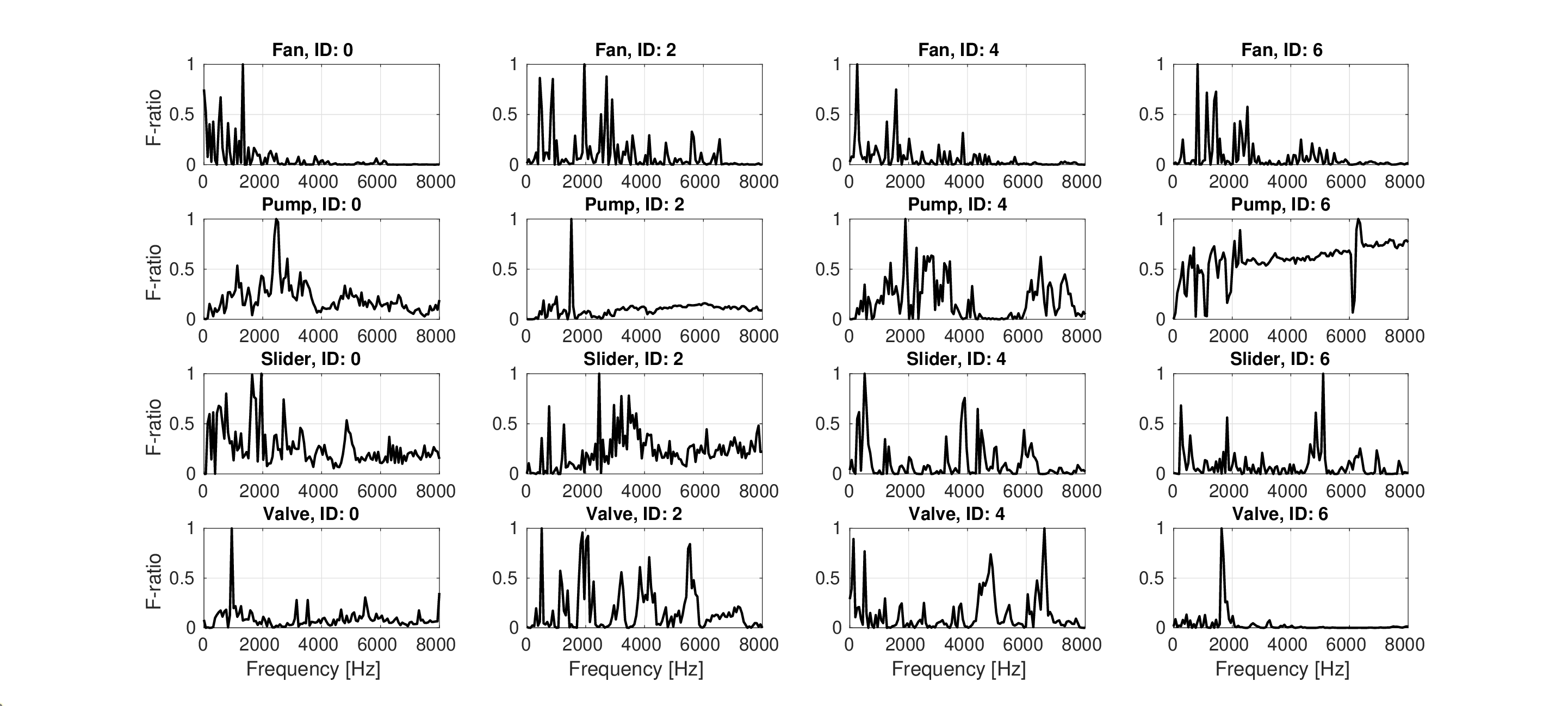}
\caption{Quantification results of frequency importance for different machines and IDs using F-ratio. Data with SNR = 6 were used for calculations.}
\label{fig:fratios}
\end{figure*}

\begin{table}[t]
\centering
\caption{Statistics of normal and abnormal sound examples in MIMII dataset for various machine types and their different IDs at given SNR level. Dataset consists of three parallel subsets, available at 6, 0, and -6 dB SNR levels.}
\label{tab:dataset}
\begin{tabular}{lccc}
\hline\hline
\multicolumn{1}{c}{\textbf{Machine}} & \textbf{ID} & \textbf{\begin{tabular}[c]{@{}c@{}}Number of \\ normal sounds\end{tabular}} & \textbf{\begin{tabular}[c]{@{}c@{}}Number of \\ abnormal sounds\end{tabular}} \\ \hline
\multirow{4}{*}{fan}                 & 0           & 1011                                                                        & 407                                                                           \\
                                     & 2           & 1016                                                                        & 359                                                                           \\
                                     & 4           & 1033                                                                        & 348                                                                           \\
                                     & 6           & 1015                                                                        & 361                                                                           \\ \hline
\multirow{4}{*}{pump}                & 0           & 1006                                                                        & 143                                                                           \\
                                     & 2           & 1005                                                                        & 111                                                                           \\
                                     & 4           & 702                                                                         & 100                                                                           \\
                                     & 6           & 1036                                                                        & 102                                                                           \\ \hline
\multirow{4}{*}{slider}              & 0           & 991                                                                         & 119                                                                           \\
                                     & 2           & 708                                                                         & 120                                                                           \\
                                     & 4           & 1000                                                                        & 120                                                                           \\
                                     & 6           & 992                                                                         & 120                                                                           \\ \hline
\multirow{4}{*}{valve}               & 0           & 1068                                                                        & 356                                                                           \\
                                     & 2           & 1068                                                                        & 267                                                                           \\
                                     & 4           & 534                                                                         & 178                                                                           \\
                                     & 6           & 534                                                                         & 89                                                                            \\ \hline
Total                                &             & 14719                                                                       & 3300                                                                          \\ \hline\hline
\end{tabular}
\end{table}

\begin{figure*}[t]
\centering
\includegraphics[width=\textwidth,trim=100 0 100 0]{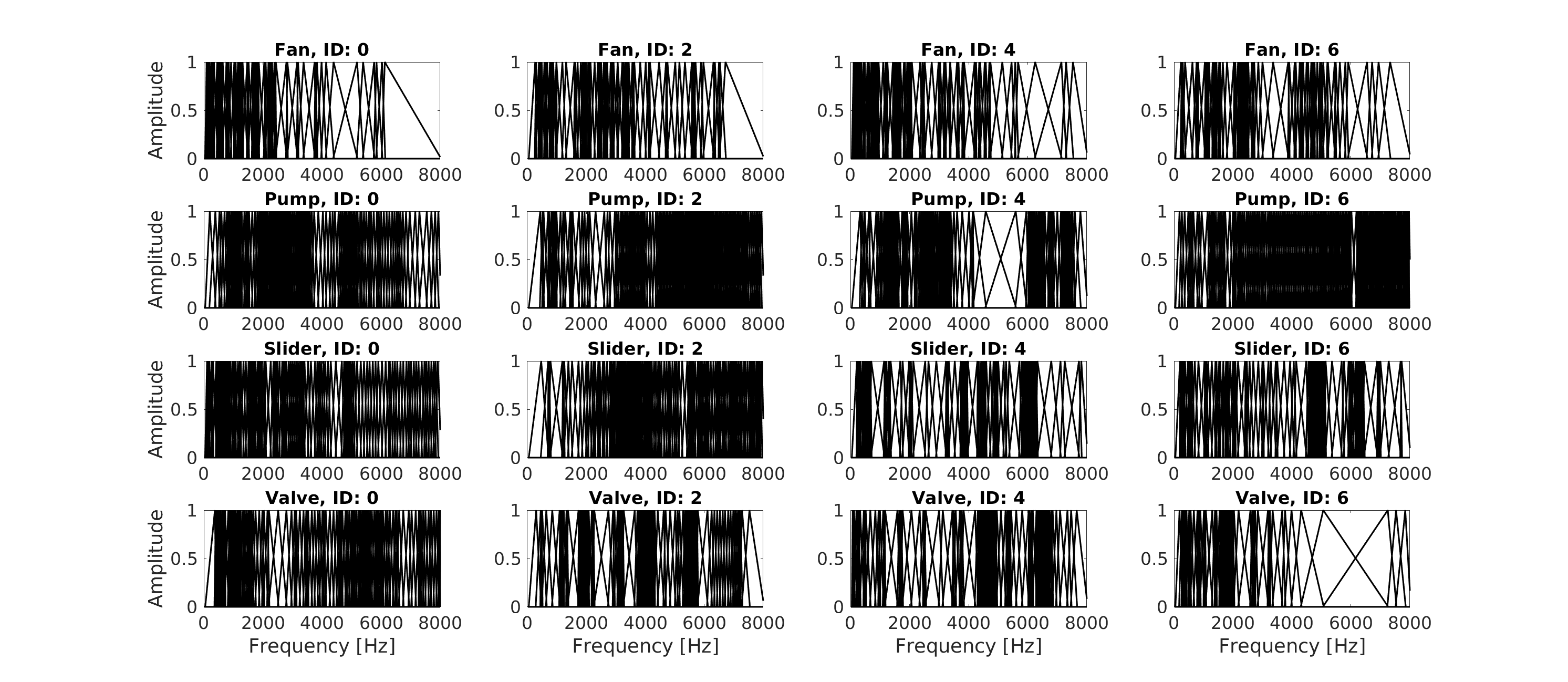}
\caption{The designed data-driven filterbanks for each machine and ID.}
\label{fig:filterbanks}
\end{figure*}

\section{Database Statistics}
\label{sec:data}
The MIMII dataset \cite{purohit2019mimii} comprises normal and abnormal sounds from four types of industrial machines: fans, pumps, sliders, and valves. Each type of machine includes multiple individual models identified by specific IDs. For instance, fan machines have four different IDs: 0, 2, 4, and 6. Each sound example in the dataset is 10 seconds long, recorded at a 16-kHz sampling rate using 8 microphones.

The MIMII dataset is publicly available with recordings at three different SNRs: -6, 0, and 6 dB. Real factory noise was recorded in multiple factories and added to the original machine sounds to create these different SNR conditions. Table \ref{tab:dataset} presents the data distribution for each machine type and machine ID at the specified SNRs. Notably, there are three subsets of sound recordings, each corresponding to a different SNR level.


Normal and abnormal sounds were recorded in a reverberant environment, capturing the typical operating conditions of these machines. The background noise, recorded in real factories, was mixed with the machine sounds to simulate realistic conditions. The dataset includes a total of 14,719 sound files for normal conditions and 3,300 for abnormal conditions, as detailed in Table \ref{tab:dataset}.

We seleectd the MIMII dataset for our experiments due to its comprehensive coverage of machine types and realistic noise conditions. Fans exhibit unbalanced or clogging problems in abnormal situations, pumps encounter leakage or clogging problems, sliders suffer from rail damage or lack of grease, and valves encounter various contaminations. This dataset provides a robust foundation for developing and testing anomaly detection methods in industrial settings.

\section{Experimental Setup and Evaluation Metric}
\label{sec:ExpermentalSetup}
To extract the LMS feature, $10$-s audio clips were first split into different frames with frame lengths of $64$ ms and hop lengths of $32$ ms. The Mel spectrogram feature was then extracted with the following parameters: $n\_fft$=1024, $hop\_length$=512, $num\_filters$=128, and $power=2.0$. We extracted the LNS and LGS features using the same configuration but different types of filterbanks compared with the LMS feature. Five consecutive frames with a sliding window were concatenated into one feature vector with a dimension of 640 and fed into the detector. For example, we assumes that the input signal is $X=\{X_t\}_{t=1}^T$ where $X_t\in\mathbb{R}^M$, and $M$ and $T$ are the number of Mel-filters and time-frames, respectively. Then, the acoustic feature at $t$ was obtained by concatenating consecutive frames of the feature as $\psi_t\in\mathbb{R}^D$, where $D=P\times M$, $P=5$, $M=128$ and $D=640$. For the modulation representations, the dimension $D=P\times M\times K$, where $K$ is the number of modulation channels. $K$ is equal to 6, 6, and 12 for SM, TM and STM, respectively. The reconstruction
error is calculated as
\begin{equation}
E(X)=\frac{1}{DT}\sum_{t=1}^T \parallel \psi_t-r(\psi_t)\parallel_2^2,
\end{equation}
where $r(\psi_t)$ is the vector reconstructed with the AE model, and $\parallel \cdot\parallel_2$ is $\mathcal{L}_2$ norm.

In our implementation, three variations of the AE model were explored, each differing in the number of neurons in the fully connected layers: 64, 128, and 256. These variations enable us to assess the impact of model complexity on detection performance. Detailed information on the layer structure for each variation is provided in Table \ref{table:models}.

Modulation filters include a low-pass filter with a cut-off frequency of 2 Hz and band-pass filters with cut-off frequencies of [2, 4], [4, 8], [8, 16], [16, 32], and [32, 64] Hz. All filters were designed with the Butterworth infinite impulse response (IIR) filter.

We used the area under the receiving operating characteristic curve (AUC) as the evaluation metric to assess the performance of proposed features and representations. The AUC represents the degree or measure of separability achieved with the AE model: an AUC of 1.0 indicates perfect classification, while an AUC of 0.5 suggests no discriminative power, equivalent to random guessing.

\begin{table}[]
\centering
\caption{Parameters of different models}
\label{table:models}
\begin{tabular}{cc}
\hline\hline
\textbf{Model index} & \textbf{Layer structure}                         \\ \hline
1                    & {[}64, 64, 64, 64, 8, 64, 64, 64, 64{]}          \\
2                    & {[}128, 128, 128, 128, 16, 128, 128, 128, 128{]} \\
3                    & {[}256, 256, 256, 256, 32, 256, 256, 256, 256{]} \\ \hline\hline
\end{tabular}
\end{table}

\section{Evaluation Results}
\label{sec:results}
Timbre features \cite{zwicker2013psychoacoustics}, such as boominess, brightness, depth, roughness, and sharpness, are well-established in audio analysis for their effectiveness in capturing the tonal quality and texture of sounds. These features are crucial for differentiating between various sound sources and understanding the auditory characteristics. The modulation representations, however, are highly effective in capturing dynamic variations in both time and frequency domains, including not only the qualities reflected by timbre features but also additional discriminative information such as temporal rhythm and spectral changes \cite{ota2023anomalous}. We compared the performance of timbre features with the proposed features and representations. By comparing these two feature sets, the aim was to evaluate whether the STM representations offer a significant advantage over timbre features in distinguishing machine anomalies.

In this section, we present and analyze the experimental results on machine ASD using the MIMII dataset. We evaluated the performance of proposed LNS feature in comparison with the timbral, LMS, and LGS features. We also discuss the effectiveness of the STM representations derived from LMS, LGS, and LNS features under SNRs of -6 dB, 0 dB, and 6 dB. We investigated the impact of different model configurations on the performance of STM representations derived from the LNS feature. Finally, we compared the performance of proposed representations with state-of-the-art methods that use the MIMII dataset, demonstrating the efficacy of our approach.

\begin{figure*}[t]
\centering
\includegraphics[width=7in]{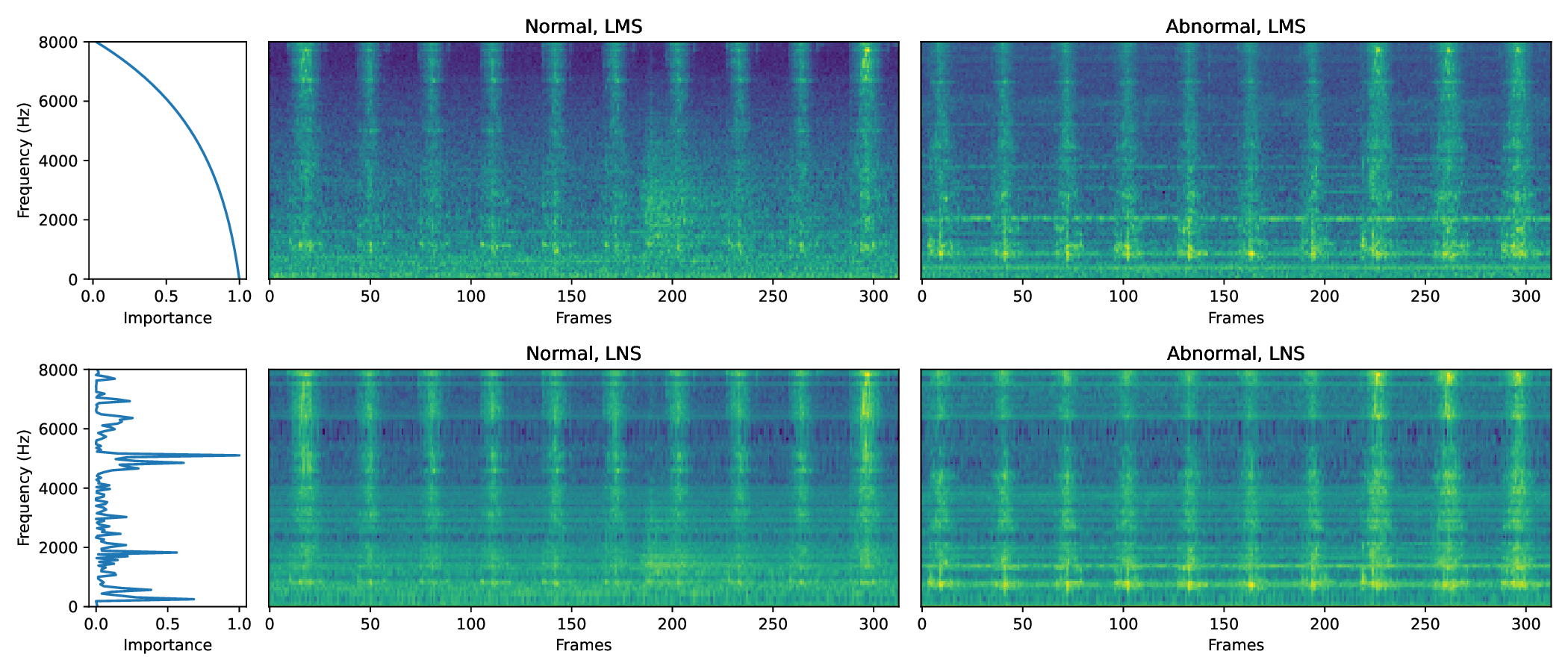}
\caption{Comparison of LMS and LNS features using normal and abnormal audios emitted from slider ID 6 with SNR = 6 dB.}
\label{fig:spectrogram}
\end{figure*}

\subsection{Results of Frequency Quantification}
Figure \ref{fig:fratios} presents the quantification results of frequency importance for different machines and IDs using the F-ratio. Training data with SNR = 6 dB were used for the calculations and the frequency importance was normalized from 0 to 1.

The MFB and GFB are designed according to the pitch perception and auditory perception of the human ear, respectively, and traditionally emphasize lower-frequency regions. However, the F-ratio results in Fig. \ref{fig:fratios} reveal that discriminative information between normal and anomalous sounds is distributed non-uniformly across the frequency spectrum. For some machines and their specific IDs, higher-frequency regions contain significant discriminative information. For instance, in the case of the Pump ID 4, slider ID 6, and Valve ID 4, there were notable peaks in the mid- to high- frequency regions, indicating that these frequencies are important for distinguishing between normal and anomalous sounds. This contrasts with the general MFB tendency and highlights the need for machine-specific frequency analysis.

Similar machines exhibited consistent trends in frequency band importance. For example, the fans (Fan IDs 0, 2, 4, and 6) consistently showed higher importance in the lower-frequency regions, though the exact distribution varies slightly across different IDs. The designed machine-specific NUFBs are illustrated in Fig. \ref{fig:filterbanks}. Narrower bandwidth and higher filter distribution density appear in the frequency regions that exhibit higher F-ratios.

Figure \ref{fig:spectrogram} presents a comparative analysis of the LMS and proposed LNS feature using normal and abnormal audio signals emitted from slider ID 6 with an SNR of 6 dB. The figure illustrates how the LNS feature, designed based on the quantification results, enhance the resolution and discriminative power of frequency regions with higher F-ratios. In the LMS feature, the frequency importance decreases from lower to higher frequencies, resulting in a broader and smoother frequency representation. In contrast, the LNS feature focus on regions with higher F-ratios, leading to finer granularity and clearer representation of these critical frequencies. The differences between normal and abnormal sounds are more distinct in the LNS feature. This is especially visible in the frequency regions, where the F-ratios are higher. For instance, at frequencies around 100 and 5,000 Hz, the LNS feature reveal distinct patterns that differentiate normal and abnormal conditions more effectively than the LMS.

\begin{table*}[htbp]
\centering
\caption{Results comparison of timbre, LMS, LGS, and the proposed LNS features.}
\label{tab:unfb}
\begin{tabular}{lccccccccccc}
\hline\hline
\rowcolor[HTML]{FFFFFF} 
\multicolumn{1}{c}{\cellcolor[HTML]{FFFFFF}\textbf{Machine}} & \textbf{ID}               & \textbf{Boominess}            & \textbf{Brightness}           & \textbf{Depth}                & \textbf{Roughness}            & \textbf{Sharpness}            & \textbf{AS}                   & \textbf{APF}                  & \textbf{LMS}                  & \textbf{LGS}                  & \textbf{LNS}                 \\ \hline
\cellcolor[HTML]{FFFFFF}                                     & \cellcolor[HTML]{FFFFFF}0 & \cellcolor[HTML]{C3C3C3}0.683 & \cellcolor[HTML]{C9C9C9}0.656 & \cellcolor[HTML]{ACACAC}0.795 & \cellcolor[HTML]{C3C3C3}0.687 & \cellcolor[HTML]{DFDFDF}0.551 & \cellcolor[HTML]{DEDEDE}0.557 & \cellcolor[HTML]{BABABA}0.727 & \cellcolor[HTML]{B2B2B2}0.764 & \cellcolor[HTML]{AEAEAE}0.784 & \cellcolor[HTML]{9D9D9D}0.866 \\
\cellcolor[HTML]{FFFFFF}                                     & \cellcolor[HTML]{FFFFFF}2 & \cellcolor[HTML]{D6D6D6}0.595 & \cellcolor[HTML]{888888}0.965 & \cellcolor[HTML]{CACACA}0.652 & \cellcolor[HTML]{8A8A8A}0.957 & \cellcolor[HTML]{A4A4A4}0.832 & \cellcolor[HTML]{9B9B9B}0.874 & \cellcolor[HTML]{C3C3C3}0.687 & \cellcolor[HTML]{828282}0.993 & \cellcolor[HTML]{898989}0.962 & \cellcolor[HTML]{828282}0.991 \\
\cellcolor[HTML]{FFFFFF}                                     & \cellcolor[HTML]{FFFFFF}4 & \cellcolor[HTML]{ADADAD}0.791 & \cellcolor[HTML]{AAAAAA}0.802 & \cellcolor[HTML]{B3B3B3}0.763 & \cellcolor[HTML]{A2A2A2}0.842 & \cellcolor[HTML]{CDCDCD}0.637 & \cellcolor[HTML]{C0C0C0}0.699 & \cellcolor[HTML]{D1D1D1}0.620 & \cellcolor[HTML]{949494}0.908 & \cellcolor[HTML]{878787}0.967 & \cellcolor[HTML]{8B8B8B}0.948 \\
\multirow{-4}{*}{\cellcolor[HTML]{FFFFFF}Fan}                & \cellcolor[HTML]{FFFFFF}6 & \cellcolor[HTML]{C3C3C3}0.683 & \cellcolor[HTML]{989898}0.890 & \cellcolor[HTML]{BCBCBC}0.718 & \cellcolor[HTML]{AFAFAF}0.780 & \cellcolor[HTML]{A9A9A9}0.810 & \cellcolor[HTML]{898989}0.960 & \cellcolor[HTML]{909090}0.925 & \cellcolor[HTML]{818181}0.997 & \cellcolor[HTML]{808080}1.000 & \cellcolor[HTML]{828282}0.994 \\ \hline
\cellcolor[HTML]{FFFFFF}                                     & \cellcolor[HTML]{FFFFFF}0 & \cellcolor[HTML]{A4A4A4}0.831 & \cellcolor[HTML]{A7A7A7}0.816 & \cellcolor[HTML]{C0C0C0}0.701 & \cellcolor[HTML]{A9A9A9}0.807 & \cellcolor[HTML]{ADADAD}0.788 & \cellcolor[HTML]{888888}0.963 & \cellcolor[HTML]{BDBDBD}0.712 & \cellcolor[HTML]{AFAFAF}0.781 & \cellcolor[HTML]{A6A6A6}0.824 & \cellcolor[HTML]{A8A8A8}0.815 \\
\cellcolor[HTML]{FFFFFF}                                     & \cellcolor[HTML]{FFFFFF}2 & \cellcolor[HTML]{D5D5D5}0.602 & \cellcolor[HTML]{A4A4A4}0.831 & \cellcolor[HTML]{B6B6B6}0.748 & \cellcolor[HTML]{BBBBBB}0.724 & \cellcolor[HTML]{9F9F9F}0.856 & \cellcolor[HTML]{D0D0D0}0.625 & \cellcolor[HTML]{C4C4C4}0.682 & \cellcolor[HTML]{ECECEC}0.491 & \cellcolor[HTML]{F2F2F2}0.460 & \cellcolor[HTML]{CBCBCB}0.649 \\
\cellcolor[HTML]{FFFFFF}                                     & \cellcolor[HTML]{FFFFFF}4 & \cellcolor[HTML]{D1D1D1}0.619 & \cellcolor[HTML]{CECECE}0.634 & \cellcolor[HTML]{D5D5D5}0.601 & \cellcolor[HTML]{C7C7C7}0.667 & \cellcolor[HTML]{9E9E9E}0.861 & \cellcolor[HTML]{C4C4C4}0.680 & \cellcolor[HTML]{CDCDCD}0.638 & \cellcolor[HTML]{868686}0.973 & \cellcolor[HTML]{989898}0.891 & \cellcolor[HTML]{818181}0.997 \\
\multirow{-4}{*}{\cellcolor[HTML]{FFFFFF}Pump}               & \cellcolor[HTML]{FFFFFF}6 & \cellcolor[HTML]{979797}0.893 & \cellcolor[HTML]{BDBDBD}0.715 & \cellcolor[HTML]{9A9A9A}0.880 & \cellcolor[HTML]{BFBFBF}0.706 & \cellcolor[HTML]{D1D1D1}0.618 & \cellcolor[HTML]{C2C2C2}0.690 & \cellcolor[HTML]{A4A4A4}0.832 & \cellcolor[HTML]{9F9F9F}0.855 & \cellcolor[HTML]{8A8A8A}0.957 & \cellcolor[HTML]{DFDFDF}0.554 \\ \hline
\cellcolor[HTML]{FFFFFF}                                     & \cellcolor[HTML]{FFFFFF}0 & \cellcolor[HTML]{858585}0.980 & \cellcolor[HTML]{848484}0.985 & \cellcolor[HTML]{878787}0.968 & \cellcolor[HTML]{909090}0.925 & \cellcolor[HTML]{828282}0.993 & \cellcolor[HTML]{818181}0.997 & \cellcolor[HTML]{C5C5C5}0.675 & \cellcolor[HTML]{828282}0.992 & \cellcolor[HTML]{808080}1.000 & \cellcolor[HTML]{808080}1.000 \\
\cellcolor[HTML]{FFFFFF}                                     & \cellcolor[HTML]{FFFFFF}2 & \cellcolor[HTML]{C5C5C5}0.675 & \cellcolor[HTML]{AFAFAF}0.781 & \cellcolor[HTML]{A2A2A2}0.839 & \cellcolor[HTML]{949494}0.908 & \cellcolor[HTML]{ADADAD}0.788 & \cellcolor[HTML]{B6B6B6}0.745 & \cellcolor[HTML]{B8B8B8}0.738 & \cellcolor[HTML]{929292}0.919 & \cellcolor[HTML]{818181}0.998 & \cellcolor[HTML]{8D8D8D}0.939 \\
\cellcolor[HTML]{FFFFFF}                                     & \cellcolor[HTML]{FFFFFF}4 & \cellcolor[HTML]{BBBBBB}0.722 & \cellcolor[HTML]{9D9D9D}0.867 & \cellcolor[HTML]{B4B4B4}0.755 & \cellcolor[HTML]{8B8B8B}0.948 & \cellcolor[HTML]{9D9D9D}0.864 & \cellcolor[HTML]{ACACAC}0.792 & \cellcolor[HTML]{C7C7C7}0.664 & \cellcolor[HTML]{9A9A9A}0.877 & \cellcolor[HTML]{868686}0.976 & \cellcolor[HTML]{808080}1.000 \\
\multirow{-4}{*}{\cellcolor[HTML]{FFFFFF}slider}             & \cellcolor[HTML]{FFFFFF}6 & \cellcolor[HTML]{C6C6C6}0.670 & \cellcolor[HTML]{C8C8C8}0.662 & \cellcolor[HTML]{C7C7C7}0.666 & \cellcolor[HTML]{B4B4B4}0.756 & \cellcolor[HTML]{D1D1D1}0.621 & \cellcolor[HTML]{CACACA}0.650 & \cellcolor[HTML]{CFCFCF}0.627 & \cellcolor[HTML]{C3C3C3}0.685 & \cellcolor[HTML]{ADADAD}0.789 & \cellcolor[HTML]{848484}0.982 \\ \hline
\cellcolor[HTML]{FFFFFF}                                     & \cellcolor[HTML]{FFFFFF}0 & \cellcolor[HTML]{949494}0.908 & \cellcolor[HTML]{EAEAEA}0.498 & \cellcolor[HTML]{A9A9A9}0.809 & \cellcolor[HTML]{9C9C9C}0.871 & \cellcolor[HTML]{969696}0.900 & \cellcolor[HTML]{838383}0.990 & \cellcolor[HTML]{D1D1D1}0.617 & \cellcolor[HTML]{BEBEBE}0.708 & \cellcolor[HTML]{9B9B9B}0.875 & \cellcolor[HTML]{838383}0.988 \\
\cellcolor[HTML]{FFFFFF}                                     & \cellcolor[HTML]{FFFFFF}2 & \cellcolor[HTML]{C6C6C6}0.669 & \cellcolor[HTML]{CBCBCB}0.647 & \cellcolor[HTML]{C8C8C8}0.662 & \cellcolor[HTML]{C3C3C3}0.687 & \cellcolor[HTML]{D0D0D0}0.623 & \cellcolor[HTML]{BFBFBF}0.704 & \cellcolor[HTML]{CACACA}0.650 & \cellcolor[HTML]{CDCDCD}0.638 & \cellcolor[HTML]{BBBBBB}0.723 & \cellcolor[HTML]{9C9C9C}0.869 \\
\cellcolor[HTML]{FFFFFF}                                     & \cellcolor[HTML]{FFFFFF}4 & \cellcolor[HTML]{C4C4C4}0.680 & \cellcolor[HTML]{CECECE}0.632 & \cellcolor[HTML]{CECECE}0.631 & \cellcolor[HTML]{B6B6B6}0.748 & \cellcolor[HTML]{CECECE}0.631 & \cellcolor[HTML]{A9A9A9}0.807 & \cellcolor[HTML]{9B9B9B}0.873 & \cellcolor[HTML]{CECECE}0.632 & \cellcolor[HTML]{ADADAD}0.790 & \cellcolor[HTML]{9F9F9F}0.857 \\
\multirow{-4}{*}{\cellcolor[HTML]{FFFFFF}Valve}              & \cellcolor[HTML]{FFFFFF}6 & \cellcolor[HTML]{CACACA}0.653 & \cellcolor[HTML]{B9B9B9}0.731 & \cellcolor[HTML]{CCCCCC}0.641 & \cellcolor[HTML]{C5C5C5}0.676 & \cellcolor[HTML]{BABABA}0.726 & \cellcolor[HTML]{C5C5C5}0.674 & \cellcolor[HTML]{D3D3D3}0.607 & \cellcolor[HTML]{BDBDBD}0.715 & \cellcolor[HTML]{A7A7A7}0.816 & \cellcolor[HTML]{CACACA}0.653 \\ \hline
\multicolumn{12}{l}{Average on each machine}                                                                                                                                                                                                                                                                                                                                                                             \\ \hline
\multicolumn{2}{l}{\cellcolor[HTML]{FFFFFF}Fan}                                          & \cellcolor[HTML]{DFDFDF}0.688 & \cellcolor[HTML]{B2B2B2}0.828 & \cellcolor[HTML]{D1D1D1}0.732 & \cellcolor[HTML]{B5B5B5}0.817 & \cellcolor[HTML]{D9D9D9}0.707 & \cellcolor[HTML]{C4C4C4}0.773 & \cellcolor[HTML]{CECECE}0.740 & \cellcolor[HTML]{959595}0.916 & \cellcolor[HTML]{919191}0.928 & \cellcolor[HTML]{8A8A8A}0.950 \\
\multicolumn{2}{l}{\cellcolor[HTML]{FFFFFF}Pump}                                         & \cellcolor[HTML]{CFCFCF}0.736 & \cellcolor[HTML]{CBCBCB}0.749 & \cellcolor[HTML]{D0D0D0}0.733 & \cellcolor[HTML]{D3D3D3}0.726 & \cellcolor[HTML]{C1C1C1}0.781 & \cellcolor[HTML]{CECECE}0.739 & \cellcolor[HTML]{D6D6D6}0.716 & \cellcolor[HTML]{C3C3C3}0.775 & \cellcolor[HTML]{C0C0C0}0.783 & \cellcolor[HTML]{CACACA}0.754 \\
\multicolumn{2}{l}{\cellcolor[HTML]{FFFFFF}Slider}                                       & \cellcolor[HTML]{C7C7C7}0.762 & \cellcolor[HTML]{B3B3B3}0.824 & \cellcolor[HTML]{B8B8B8}0.807 & \cellcolor[HTML]{A0A0A0}0.884 & \cellcolor[HTML]{B5B5B5}0.817 & \cellcolor[HTML]{BCBCBC}0.796 & \cellcolor[HTML]{E3E3E3}0.676 & \cellcolor[HTML]{A5A5A5}0.868 & \cellcolor[HTML]{8D8D8D}0.941 & \cellcolor[HTML]{808080}0.980 \\
\multicolumn{2}{l}{\cellcolor[HTML]{FFFFFF}Valve}                                        & \cellcolor[HTML]{D2D2D2}0.727 & \cellcolor[HTML]{F2F2F2}0.627 & \cellcolor[HTML]{DFDFDF}0.686 & \cellcolor[HTML]{CCCCCC}0.745 & \cellcolor[HTML]{D4D4D4}0.720 & \cellcolor[HTML]{BDBDBD}0.794 & \cellcolor[HTML]{DFDFDF}0.687 & \cellcolor[HTML]{E4E4E4}0.673 & \cellcolor[HTML]{BABABA}0.801 & \cellcolor[HTML]{ADADAD}0.842 \\
\multicolumn{2}{l}{\cellcolor[HTML]{FFFFFF}Average in total}                             & \cellcolor[HTML]{D2D2D2}0.728 & \cellcolor[HTML]{C9C9C9}0.757 & \cellcolor[HTML]{CECECE}0.739 & \cellcolor[HTML]{BDBDBD}0.793 & \cellcolor[HTML]{C9C9C9}0.756 & \cellcolor[HTML]{C3C3C3}0.775 & \cellcolor[HTML]{D9D9D9}0.705 & \cellcolor[HTML]{B8B8B8}0.808 & \cellcolor[HTML]{A6A6A6}0.863 & \cellcolor[HTML]{A0A0A0}0.881 \\ \hline\hline
\end{tabular}
\end{table*}

\subsection{Results of proposed LNS feature}
Table \ref{tab:unfb} presents the comparison of various features, including timbral, LMS, LGS, and LNS features, for four types of machines at an SNR of 6 dB. Timbral features include boominess, brightness, depth, roughness, sharpness, amplified shimmer (AS), and amplified predominant frequency (APF), as reported in a previous study \cite{ota2023anomalous}. The results of the LNS feature are proposed in this paper.

The LNS feature consistently outperformed the timbral features in all machine types except the pump. For instance, in the case of the fan, the LNS feature achieved an average score of 0.950, significantly higher than the best timbral feature (brightness), which has an average score of 0.828. Similarly, for the slider, the LNS feature attained an average score of 0.980, compared with the highest timbral feature score of 0.884 (roughness). The LNS feature also demonstrated superior performance compared with the LMS and LGS features. For the slider, the LNS feature achieved an average score of 0.980, while the LMS and LGS features reached 0.868 and 0.941, respectively. For the valve machine, the LNS feature showed a marked improvement with a score of 0.842 compared with 0.673 and 0.801 from the LMS and LGS features, respectively.

The slider showed the most significant improvement with LNS feature, with perfect scores for IDs 0 and 4, indicating its effectiveness in capturing anomaly-related patterns. The fan and valve also benefited notably from the LNS feature, though the improvement was less pronounced than for the slider. On average, the LNS feature achieved the highest scores across all machines with a total average score of 0.881, compared with the average scores of 0.808 and 0.863 for the LMS and LGS features, respectively, significantly higher than the average scores of the individual timbral features.

While the overall performance of the LNS feature was superior, there were instances in which the performance was suboptimal. Notable examples include pump ID 6 and valve ID 6. This problem can be attributed to the independent calculation method of frequency band importance while using the F-ratio. Consequently, unique operational characteristics or noise patterns of pump ID 6 and valve ID 6 may not be effectively captured with the LNS feature. 

Unlike the timbral features, which are fixed and may not effectively capture relevant information across different types of sounds, the LNS feature were designed to adaptively emphasize important frequency regions, thus capturing more discriminative features to the distinction between normal and anomalous sounds, as demonstrated by the consistent improvement in detection scores.

\subsection{Results of Spectral-temporal Modulation Analysis}

Table \ref{tab:stmall} presents the averaged results for different modulation representations derived from the LMS, LGS, and LNS features in four types of machines under varying SNR conditions (-6, 0, and 6 dB). Despite being calculated from data with an SNR of 6 dB, the LNS feature performed well across all SNR levels, consistently outperforming the LMS features in average AUC. For example, at -6 dB SNR, the average AUC for the LNS feature was 0.702, while that for the LMS feature was 0.662. This trend continued at 0 and 6 dB SNR levels, with the LNS feature achieving average AUC scores of 0.785 and 0.881, respectively.

The STM representation derived from the LNS feature outperformed those derived from the LMS and LGS features across all SNR conditions on average. For example, at -6 dB SNR, STM (LNS) achieved an AUC of 0.702 compared to 0.624 and 0.657 of STM (LMS) and STM (LGS), respectively. The TM derived from the LNS feature showed a significant improvement over the LMS, particularly for the fan, pump, and slider. At 6 dB SNR, TM (LNS) achieved AUC scores of 0.974, 0.945, and 0.930, respectively. The SM derived from the LNS feature performed well for the valve. At -6 dB SNR, SM (LNS) achieved an AUC of 0.737, compared to 0.434 and 0.505 for SM (LMS) and SM (LGS), respectively. 

Although STM combine the advantages of SM and TM, it does not always produce the best performance for all types of machines. This is evident in the valve at 0 dB SNR, where STM (LNS) achieved an AUC of 0.528, which is lower than both SM (0.757) and TM (0.559). This may be due to the increased complexity of STM representations, which could lead to insufficient capture of anomaly patterns when using a simple AE model.

Overall, the results indicate that the proposed LNS feature, particularly in the TM and STM representations, significantly enhance anomaly detection performance across various machine types and SNR levels. However, certain types of machines, such as valves, require further optimization to improve detection accuracy. A more detailed discussion on why the effectiveness of modulation representations varies across different machines is provided in Section \ref{sec:Discussion}.

\begin{table*}[]
\centering
\caption{Performances of SM, TM, and STM representations derived from LMS, LGS, and LNS features. The results are evaluated in terms of AUC. All results were calculated by averaging machine IDs.}
\label{tab:stmall}
\begin{tabular}{lcl|ccc|ccc|ccc|ccc}
\hline\hline
\multicolumn{1}{c}{\textbf{SNR}} & \multicolumn{2}{c|}{\textbf{Machine}} & \textbf{LMS} & \textbf{LGS}   & \textbf{LNS}   & \textbf{\begin{tabular}[c]{@{}c@{}}SM \\ (LMS)\end{tabular}} & \textbf{\begin{tabular}[c]{@{}c@{}}SM \\ (LGS)\end{tabular}} & \textbf{\begin{tabular}[c]{@{}c@{}}SM \\ (LNS)\end{tabular}} & \textbf{\begin{tabular}[c]{@{}c@{}}TM \\ (LMS)\end{tabular}} & \textbf{\begin{tabular}[c]{@{}c@{}}TM \\ (LGS)\end{tabular}} & \textbf{\begin{tabular}[c]{@{}c@{}}TM \\ (LNS)\end{tabular}} & \textbf{\begin{tabular}[c]{@{}c@{}}STM \\ (LMS)\end{tabular}} & \textbf{\begin{tabular}[c]{@{}c@{}}STM \\ (LGS)\end{tabular}} & \textbf{\begin{tabular}[c]{@{}c@{}}STM \\ (LNS)\end{tabular}} \\ \hline
\multirow{5}{*}{-6 dB}           & \multicolumn{2}{c|}{Fan}              & 0.672        & 0.641          & \textbf{0.716} & 0.510                                                        & 0.555                                                        & 0.540                                                        & 0.661                                                        & 0.762                                                        & 0.698                                                        & 0.691                                                         & 0.675                                                         & \textbf{0.716}                                                \\
                                 & \multicolumn{2}{c|}{Pump}             & 0.675        & 0.697          & 0.775          & 0.580                                                        & 0.608                                                        & 0.590                                                        & 0.726                                                        & 0.756                                                        & \textbf{0.784}                                               & 0.652                                                         & 0.758                                                         & 0.775                                                         \\
                                 & \multicolumn{2}{c|}{Slider}           & 0.725        & 0.807          & \textbf{0.813} & 0.748                                                        & 0.673                                                        & 0.798                                                        & 0.665                                                        & 0.723                                                        & 0.808                                                        & 0.656                                                         & 0.721                                                         & \textbf{0.813}                                                \\
                                 & \multicolumn{2}{c|}{Valve}            & 0.578        & 0.682          & 0.505          & 0.434                                                        & 0.505                                                        & \textbf{0.737}                                               & 0.510                                                        & 0.506                                                        & 0.527                                                        & 0.496                                                         & 0.473                                                         & 0.505                                                         \\
                                 & \multicolumn{2}{c|}{Average in total} & 0.662        & \textbf{0.707} & 0.702          & 0.568                                                        & 0.585                                                        & 0.666                                                        & 0.640                                                        & 0.687                                                        & 0.704                                                        & 0.624                                                         & 0.657                                                         & 0.702                                                         \\ \hline
\multirow{5}{*}{0 dB}            & \multicolumn{2}{c|}{Fan}              & 0.807        & 0.768          & 0.862          & 0.634                                                        & 0.688                                                        & 0.646                                                        & 0.843                                                        & 0.902                                                        & 0.874                                                        & 0.842                                                         & \textbf{0.895}                                                & 0.862                                                         \\
                                 & \multicolumn{2}{c|}{Pump}             & 0.769        & 0.712          & 0.875          & 0.611                                                        & 0.660                                                        & 0.659                                                        & 0.864                                                        & 0.867                                                        & \textbf{0.900}                                               & 0.856                                                         & 0.858                                                         & 0.875                                                         \\
                                 & \multicolumn{2}{c|}{Slider}           & 0.771        & 0.884          & 0.874          & 0.814                                                        & 0.760                                                        & \textbf{0.889}                                               & 0.710                                                        & 0.763                                                        & 0.867                                                        & 0.689                                                         & 0.763                                                         & 0.874                                                         \\
                                 & \multicolumn{2}{c|}{Valve}            & 0.597        & \textbf{0.764} & 0.528          & 0.550                                                        & 0.575                                                        & 0.757                                                        & 0.517                                                        & 0.531                                                        & 0.559                                                        & 0.485                                                         & 0.489                                                         & 0.528                                                         \\
                                 & \multicolumn{2}{c|}{Average in total} & 0.736        & 0.782          & 0.785          & 0.652                                                        & 0.671                                                        & 0.737                                                        & 0.734                                                        & 0.766                                                        & \textbf{0.800}                                               & 0.718                                                         & 0.751                                                         & 0.785                                                         \\ \hline
\multirow{5}{*}{6 dB}            & \multicolumn{2}{c|}{Fan}              & 0.916        & 0.928          & 0.950          & 0.739                                                        & 0.776                                                        & 0.807                                                        & 0.934                                                        & 0.971                                                        & \textbf{0.974}                                               & 0.928                                                         & 0.966                                                         & 0.964                                                         \\
                                 & \multicolumn{2}{c|}{Pump}             & 0.775        & 0.938          & 0.754          & 0.762                                                        & 0.713                                                        & 0.691                                                        & 0.922                                                        & 0.914                                                        & 0.945                                                        & 0.767                                                         & 0.922                                                         & \textbf{0.953}                                                \\
                                 & \multicolumn{2}{c|}{Slider}           & 0.868        & 0.813          & \textbf{0.980} & 0.891                                                        & 0.819                                                        & 0.901                                                        & 0.816                                                        & 0.864                                                        & 0.930                                                        & 0.830                                                         & 0.853                                                         & 0.913                                                         \\
                                 & \multicolumn{2}{c|}{Valve}            & 0.673        & 0.794          & \textbf{0.842} & 0.548                                                        & 0.677                                                        & 0.840                                                        & 0.552                                                        & 0.555                                                        & 0.583                                                        & 0.483                                                         & 0.520                                                         & 0.563                                                         \\
                                 & \multicolumn{2}{c|}{Average in total} & 0.808        & 0.783          & \textbf{0.881} & 0.735                                                        & 0.746                                                        & 0.810                                                        & 0.806                                                        & 0.826                                                        & 0.858                                                        & 0.752                                                         & 0.815                                                         & 0.848                                                         \\ \hline\hline
\end{tabular}
\end{table*}

\subsection{Results of Different Model Settings}
Table \ref{tab:differentmodel} presents the AUC performance of TM, SM, and STM (derived from the proposed LNS feature) with three different model configurations across various types of machines. The best results for each modulation representation and machine type are in bold.

Performance can be further increased by increasing the number of neurons in fully connected layers. Specifically, for the fan and slider using TM, increasing the number of neurons from 64 to 128 led to an increase in AUC from 0.85 to 0.88 and from 0.87 to 0.91, respectively. However, increasing the number of neurons from 128 to 256 does not result in a significant performance improvement.

These machines benefit significantly from the TM and STM representations, with Models 2 and 3 providing the best results. This suggests that the temporal dynamics captured with these representations are crucial for detecting anomalies in these machines. The pump showed balanced performance across all feature types, with TM and STM representations performing similarly well. This indicates that both spectral and temporal information is important for this type of machine. The valve exhibited lower overall performance, with the SM performing better than the TM and STM representations. This suggests that anomalies in the valve may be more related to spectral characteristics than temporal dynamics.

\begin{table}[t]
\centering
\caption{AUC performance of modulation representations with different model configurations.}
\label{tab:differentmodel}
\begin{tabular}{cccccc}
\hline\hline
\textbf{Feature}     & \textbf{Model index} & \textbf{Fan}  & \textbf{Pump} & \textbf{Slider} & \textbf{Valve} \\ \hline
\multirow{3}{*}{TM}  & 1                    & 0.85          & 0.88          & 0.87            & 0.56           \\
                     & 2                    & \textbf{0.88} & \textbf{0.89} & \textbf{0.91}   & 0.57           \\
                     & 3                    & \textbf{0.88} & 0.88          & \textbf{0.91}   & \textbf{0.58}  \\ \hline
\multirow{3}{*}{SM}  & 1                    & 0.66          & \textbf{0.67} & \textbf{0.86}   & \textbf{0.78}  \\
                     & 2                    & \textbf{0.69} & 0.64          & 0.74            & 0.70           \\
                     & 3                    & 0.63          & 0.61          & 0.64            & 0.70           \\ \hline
\multirow{3}{*}{STM} & 1                    & 0.85          & 0.87          & 0.87            & 0.53           \\
                     & 2                    & \textbf{0.88} & \textbf{0.89} & \textbf{0.91}   & 0.57           \\
                     & 3                    & \textbf{0.88} & \textbf{0.89} & \textbf{0.91}   & \textbf{0.58}  \\ \hline\hline
\end{tabular}
\end{table}

\begin{table*}[]
\centering
\caption{Comparison of AUC performances of existing systems on the MIMII Dataset. Results averaged across SNR Levels: -6, 0, and 6 dB.}
\label{tab:comparewithothers}
\begin{tabular}{lccccc}
\hline\hline
\multicolumn{1}{c}{Paper} & Description                                                                                                      & Fan  & Pump & Slider & Valve \\ \hline
Ribeiro et al. (2020) \cite{ribeiro2020deep}       & \begin{tabular}[c]{@{}c@{}}LMS, CNN AE.\end{tabular}                                 & 0.67 & 0.72 & 0.92   & 0.79  \\
Muller  et al. (2020) \cite{muller2020acoustic}      & \begin{tabular}[c]{@{}c@{}}Mel spectrogram, \\ pretrained ImageNet model, density estimator.\end{tabular}     & 0.68 & 0.74 & 0.85   & 0.69  \\
Koizumi et al. (2020) \cite{koizumi2020description}  & LMS, AE, reconstruction error.               & 0.66 & 0.75 & 0.85   & 0.67  \\
Inoue et al. (2020) \cite{inoue2020detection}    & CNN, ensemble data augmentation.                                                                                 & 0.88 & 0.93 & 0.99   & 0.99  \\
Giri et al. (2020) \cite{giri2020unsupervised}     & DNN, MobileNetV2 ensemble, data augmentation.                                                                    & 0.82 & 0.87 & 0.97   & 0.97  \\
Singh et al. (2021) \cite{singh2021health}    & LMS, AE, reconstruction error.                & 0.80 & 0.80 & 0.85   & 0.65  \\
Kuo et al. (2022) \cite{kuo2022constructing}           & \begin{tabular}[c]{@{}c@{}}Discrete wavelet transform, fully connected DNN.\end{tabular}                      & 0.75 & 0.76 & 0.85   & 0.94  \\
Ding et al. (2023) \cite{ding2023machine}          & \begin{tabular}[c]{@{}c@{}}Discrete wavelet transform, \\ a lightweight CNN.\end{tabular} & 0.80 & 0.89 & 0.97   & 0.93  \\ \hline
Proposed (TM (NUFB))          & \begin{tabular}[c]{@{}c@{}}TM representation, AE, \\ reconstruction error.\end{tabular}                 & 0.88 & 0.89 & 0.91   & 0.57  \\
Proposed (SM (NUFB))          & \begin{tabular}[c]{@{}c@{}}SM representation, AE,\\  reconstruction error.\end{tabular}                 & 0.66 & 0.67 & 0.86   & 0.78  \\
Proposed (STM (NUFB))         & \begin{tabular}[c]{@{}c@{}}STM representation, AE,\\  reconstruction error.\end{tabular}                & 0.88 & 0.89 & 0.91   & 0.58  \\ \hline\hline
\end{tabular}
\end{table*}

\subsection{Performance Comparison with Existing Methods}
Table \ref{tab:comparewithothers} compares the averaged AUC performance of our proposed feature representations (TM, SM, and STM derived from LNS) with existing systems on the MIMII dataset. The results were averaged on four machine IDs (0, 2, 4, 6) and three SNR levels (-6, 0, and 6 dB).

The best performance for the fan was reported by Inoue et al. \cite{inoue2020detection} with an AUC of 0.88 using a convolutional neural network (CNN) with ensemble data augmentation. Our methods, both the TM and STM representations, achieved an AUC of 0.88, which is similar to the best-performing method. However, the SM representation showed a lower AUC of 0.66. In the case of the pump, the highest AUC of 0.93 was again reported by Inoue et al. \cite{inoue2020detection}. Our TM and STM representations achieved a high AUC of 0.89, which is competitive with but slightly lower than the best-performing method. 

For the slider, Inoue et al. \cite{inoue2020detection}, Giri et al. \cite{giri2020unsupervised}, and Ding et al. \cite{ding2023machine} reported outstanding AUCs of 0.99, 0.97, and 0.97, respectively. Our TM and STM representations achieved a high AUC of 0.91, which is close but not the highest. The SM representation had an AUC of 0.86. The valve showed the highest performance from Inoue et al. \cite{inoue2020detection} and Giri et al. \cite{giri2020unsupervised} with AUC of 0.99 and 0.97, respectively. Among the proposed representations, the best performance was from the SM representation with an AUC of 0.78.

In summary, our proposed representations performed well across the fan, pump, and slider, achieving outstanding results compared with most AE-based methods. However, they do not outperform methods that utilize complex training systems with data-augmentation techniques.

\section{Discussion}
\label{sec:Discussion}
\begin{figure*}[t]
\centering
\includegraphics[width=7in]{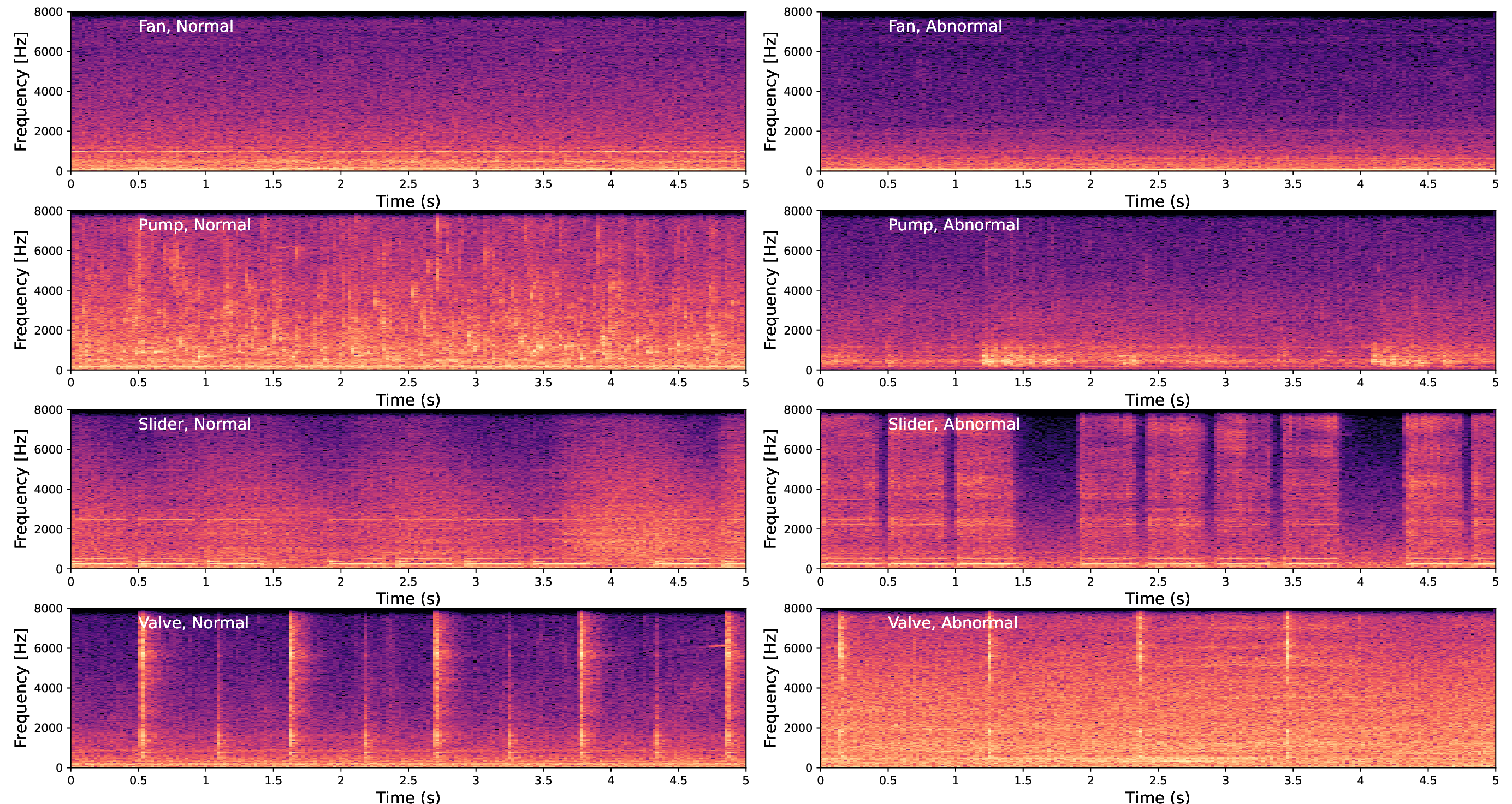}
\caption{Spectrogram comparison of normal and abnormal audio emitted from four different types of machines with SNR = 6 dB.}
\label{fig:spectros}
\end{figure*}

To explain why the proposed modulation representations, particularly TM and STM, demonstrated strong performance for the fan, pump, and slider but not for the valve, we present a comparison of spectrograms between normal and anomalous sounds for these four types of machines, as illustrated in Fig. \ref{fig:spectros}.

In the machine ASD task, modulation representations capture the dynamic changes in both the spectral and temporal domains of an audio signal. These representations are closely related to the concepts of periodicity and harmonicity, which are key features in detecting anomalies in machine sounds. Periodicity refers to the regular, repeating patterns in the time domain. Modulation representations, particularly TM, can effectively capture these periodic patterns by analyzing the rate at which the amplitude or frequency content changes over time.

Harmonicity refers to the alignment and relationship of frequencies within a sound, typically resulting in a harmonic series where the frequencies are integer multiples of a fundamental frequency. This is often seen in mechanical sounds with regular, cyclic components. SM representation are well-suited to capture harmonicity because they analyze how frequency components are organized and how they evolve over time. A normal sound might exhibit strong harmonicity, whereas an anomalous sound might introduce inharmonic components or alter the existing harmonic structure.

Industrial fans generate a continuous, stationary sound due to the constant flow of air or gas. The sound produced typically has a strong periodicity in the time domain, corresponding to the rotating blades, and harmonicity in the frequency domain due to the regular motion of the fan blades. TM and STM representations can effectively capture the periodicity and harmonicity in fan sounds. Anomalous sounds, such as changes in blade rotation speed or disruptions in airflow, can manifest as deviations from this regular pattern, which STM representation can detect. The strong periodic and harmonic structure of fan sounds makes them well-suited for detection using modulation representations, leading to improved ASD performance.

Water pumps produce a more complex sound, combining stationary components (continuous operation) with non-stationary elements (variations in flow or pressure). The sound might exhibit periodicity corresponding to the pump cycles and some harmonic structures due to the mechanical components. STM and TM representations can capture the regular, cyclical nature of a pump’s operation, as well as any anomalies that disturb this regularity. Anomalous sounds might appear as irregularities in the temporal or spectral patterns, which STM representation can detect. Since pump has both stationary and non-stationary features, modulation representations can enhance ASD performance by distinguishing between normal operation and subtle deviations caused by anomalies.

Linear slide systems generate sounds that are typically stationary but may include transient, non-stationary elements depending on the movement speed and direction. The regular movement of the slider creates a rhythmic pattern in the sound, which might have harmonicity. Modulation representations are effective in capturing the rhythmic, periodic patterns of slider sounds. Anomalies in slider operation, such as changes in speed or friction, would likely disrupt the regular sound pattern. STM representation can detect these disruptions by analyzing the deviations in both temporal and spectral modulations, resulting in improved ASD performance.

Solenoid valves produce non-stationary sounds due to repetitive opening and closing. These sounds might not exhibit strong periodicity or harmonicity because the operation is more irregular and transient. The nature of the sound could be more impulse-like, with rapid changes and less continuity. The lack of strong periodicity and harmonicity in valve sounds makes them less suitable for modulation-based analysis. STM and TM representations rely on detecting regular patterns or deviations from such patterns, but in the case of valves, the sound might not have a consistent structure for these representations to analyze effectively. As a result, the modulation representations might not capture the anomalous features as clearly as they do with the other machines, leading to poorer ASD performance.

In summary, modulation representations are effective for machines such as fans, pumps, and sliders due to the presence of periodicity and harmonicity in their normal operation sounds, which are disrupted by anomalies. However, for valves, the irregular and transient nature of the sound lacks the consistent patterns that modulation representations rely on, making them less effective for ASD in this case.

\section{Conclusion}
\label{sec:conclusion}
We quantified and visualized the importance of the frequency for different machines using a statistical method, the F-ratio. Our analysis revealed that discriminative information is distributed non-uniformly across the frequency spectrum, with certain machines and IDs showing more significant differences in the higher frequency regions. On the basis of these findings, we designed machine-specific NUFBs and proposed LNS features to emphasize frequency bands with relatively high F-ratios, enhancing the extraction of discriminative features.

We proposed STM representations derived from the proposed LNS features, with the aim of capturing variation information in the spectral and temporal domains for machine ASD. The proposed LNS feature and their modulation representations (SM, TM, and STM) were evaluated using a simple unsupervised AE model on the MIMII database with SNR levels of -6, 0, and 6 dB. The results indicated that our proposed LNS feature and modulation representations can significantly improve the performance of MFB feature. Specifically, temporal modulation proves to be effective for fans, pumps, and sliders, while spectral modulation is particularly effective for valves. Furthermore, the effectiveness of our proposed features and representations is highly competitive compared to existing methods.

Future work will focus on integrating more sophisticated neural network-based detectors to further enhance the performance of ASD systems. Furthermore, exploring other machine learning techniques and optimizing the design of filterbanks could yield further improvements, contributing to the development of more robust and efficient anomaly detection systems.




\section{Acknowledgment}
This work was supported by SCOPE Program of Ministry of Internal Affairs and Communications (Grant Number: 201605002), a Grant-in-Aid for Scientific Research (Grant number: 20H04207), a Grant-in-Aid for Scientific Research (Grant number: 21H03463), and the Fund for the Promotion of Joint International Research (Fostering Joint International Research (B))(20KK0233).

\bibliographystyle{IEEEtran}
\bibliography{refs}

\end{document}